\documentclass{ws-procs9x6}
\pdfoutput=1

\usepackage{subfigure}

\usepackage[nolists,nomarkers]{endfloat}

\begin{document}

\title{NetzCope: A Tool for Displaying and Analyzing Complex Networks}

\author{Michael J. Barber}
\address{Department of Foresight and Policy Development, Austrian Institute of Technology}

\author{Ludwig Streit}
\address{BiBoS, Univ. Bielefeld and CCM, Univ. da Madeira}

\author{Oleg Strogan}
\address{Kyiv, NaUKMA, Faculty of Informatics}

\begin{abstract}
Networks are a natural and popular mechanism for
the representation and investigation of a broad class of systems.
But extracting information from a network can present significant
challenges. We present NetzCope, a software application for the display
and analysis of networks. Its key features include the visualization
of networks in two or three dimensions, the organization of vertices to reveal
structural similarity, and the detection and visualization of network
communities by modularity maximization.
\end{abstract}


\bodymatter

\section{Introduction}

Networks describe the structure and dynamics of relations between objects or
agents: in short, networks are everywhere. In biology they have become a
prominent subject of research, in engineering network concepts have become
ever more important---from Kirchhoff's laws in the 19th century to
optimizing the design of a microchip to-day---and in the socio-economic
field ``globalization'' is paradigmatic for
their dominance. A short list of examples may serve to underscore this point:
\begin{description}
\item[Neural Networks] The well-studied nematode \textit{Caenorhabditis elegans}
has a neural network of some 300 neurons with some 7000 connections
between them.
\item[Metabolic Networks] metabolic processes in the cell.
\item[Protein interaction networks] physical interactions between an organism's proteins.
\item[Transcriptional networks] regulatory interactions between different genes.
\item[Food webs] Who eats whom?
\item[Sexual relations and infections] AIDS epidemiological models
\item[Pollination networks] plants and their pollinators
\item[Electric networks] stability of power grids
\item[Electronic networks, computer chips] computing speed
\item[Airline networks] service efficiency
\item[Internet] search engines
\item[Linguistics] words linked by co-occurrence, language families
\item[Social networks] identification of central players, gate keepers,
\item[Collaborations] actors, authors, research labs, \ldots
\end{description}
Typically these networks exhibits considerable complexity and more often
than not their structure is far from transparent. 

In this work, we present \emph{NetzCope,} a software application
for the display and analysis of complex networks. NetzCope is a
general purpose tool for investigating networks, allowing the user to
interactively explore networks, especially with regard to visualizing the
most important relationships in the network.

\subsection{Bipartite Networks}

The NetzCope software was originally developed to find and display the
structure hidden in long lists of tens of thousands of collaborative
research projects sponsored by the European Union. Said networks are \emph{%
bipartite}, with links always connecting members of two different sets. Some
examples:

\begin{description}
\item[Regulatory networks] transcription factors and target genes
\item[Economic networks] Financial centers and multinationals, firms and
board members
\item[Collaboration networks] actors and films in which they appeared together,
laboratories and joint research projects, scientists and joint publications,
\ldots
\end{description}

\subsection{Modularity}

Of particular interest to the exploration of the network of EU-funded
projects---and, indeed, to networks in general---is any possible modular
structure of the network. Quoting from an article on network biology:
\begin{quote}
Cellular functions are likely to be carried out in a highly modular
manner. In general, modularity refers to a group of physically or
functionally linked molecules (nodes) that work together to achieve a
(relatively) distinct function. Modules are seen in many systems, for
example, circles of friends in social networks or websites that are devoted
to similar topics on the World Wide Web. Similarly, in many complex
engineered systems, from a modern aircraft to a computer chip, a highly
modular structure is a fundamental design attribute. Biology is full of
examples of modularity\ldots (Barab\'{a}si and Oltvai\cite{Barabasi2004Network})
\end{quote}
To reach a more precise understanding, in order are a few words on basic concepts of graph
theory, the mathematical formulation of networks.

\section{A Few Words on Graphs}

Let
\begin{itemize}
\item V be a set (\emph{vertices})
\item E be a set of vertex pairs from $V\times V$ (\emph{edges}).
\end{itemize}
The pair $G=(V,E)$ is called a graph. Given a partition $V=V_{1}+V_{2}$. If
there are no edges between pairs of points within either $V_{i}\ $, then G
is called \emph{bipartite}.

The number of edges of a vertex $v$ we call the \emph{degree}: 
\begin{equation*}
d_{v}=\sharp \left\{ \left( v,\cdot \right) \in E\right\} 
\end{equation*}%
A simple graph $G=(V,E)$ is described by an adjacency matrix indicating
whether vertices $i$ and $k$ are connected by an edge:
\begin{equation*}
a_{ik}=\left\{ 
\begin{array}{cc}
1 & i\sim k \\ 
0 & otherwise%
\end{array}%
\right. 
\end{equation*}%
The degree of vertex $k$ is then 
\begin{equation*}
d_{k}=\sum_{i}a_{ik}
\end{equation*}%
and we shall set 
\begin{equation*}
D=\mathrm{diag}\left\{ d_{1},...,d_{n}\right\} .
\end{equation*}%
The Laplacian
\begin{equation*}
L=D-A
\end{equation*}%
and the normalized Laplacian 
\begin{equation*}
{{\mathcal{L}}}=1-D^{-1/2}AD^{-1/2}
\end{equation*}%
play a central role. In particular, $-{{\mathcal{L}}}$ is (up to a similarity
transformation) the generator of a continuous time random walk on the graph,
with equal probability $1/d_{k}$ along each edge.

\subsection{How to plot a graph}

Typically a graph or network will be given simply as a list of agents
(i.e., vertices) and their relations
(i.e., edges). How would one translate such a list
into a graphical display?

To begin with, try to arrange the vertices on a straight line: put each
vertex $k$ at position $x_{k}$ such that those connected by an edge will be
as close as possible, as if connected by elastic springs. Mechanics tells us
that such an arrangement would minimize the expression 
\begin{equation*}
E=\frac{1}{2}\sum a_{ik}\left( y_{i}-y_{k}\right) ^{2}.
\end{equation*}
Neurons in the nematode \emph{C. elegans} are said to be distributed in this
fashion!

We can also write this as the scalar product 
\begin{equation*}
E=(x,(D-A)x).
\end{equation*}%
Of course the minimum $E=0$ would arise for the vector $x=e_{0}$ with 
$x_{1}=x_{2}=...x_{k}=...=1$, indeed 
\begin{equation*}
(D-A)e_{0}=(D-A)\left( 
\begin{array}{c}
1 \\ 
\vdots \\ 
1%
\end{array}%
\right) =0.
\end{equation*}%
Here all the vertices are at the same place, $x_k=1$.
Excluding this, we are
led to the next eigenvector of $L=D-A$ with
\begin{equation*}
(D-A)e_{1}=\lambda _{1}e_{1}
\end{equation*}
In practice a better ordering is achieved using $D^{1/2}f_{1}$, where $f_1$ is
the Fiedler vector, 
the eigenvector  of
the normalized Laplacian corresponding to the smallest positive eigenvalue.

\subsection{Modularity of Graphs}

\begin{quote}
A good division of a network into communities is not merely one in which
there are few edges between communities; it is one in which there are fewer
than expected edges between communities. (M. E. J. Newman\cite{New:2006}).
\end{quote}

A popular measure of the quality of such a division or decomposition is 
the \emph{modularity}\cite{NewGir:2004}. Modularity is---up to a
normalization constant---the number of edges within communities $c$ minus
those for a null model: 
\begin{equation*}
Q\equiv \frac{1}{2|E|}\sum_{c}\sum_{i,j\in c}(A_{ij}-P_{ij}),
\end{equation*}%
where $|E|$ is the number of edges or links, and 
\begin{equation*}
P_{ij}\equiv \frac{d_{i}d_{j}}{2|E|}
\end{equation*}%
corresponds to a random graph model with a fixed set of vertices and the
constraint that on average they should reproduce a given distribution of
vertex degrees $d_{i}$\cite{ChuLu:2002}.

In empirical investigations, modularity values above roughly 0.3 are
indicative of a partitioning of the network vertices showing significant
modular structure. Modularity close to one would correspond to a near
perfect decomposition of the network into loosely interconnected communities.

The goal now is to find a division of the vertices into communities such
that the modularity $Q$ is maximal. An exhaustive search for such a
decomposition is out of the question: even for moderately large graphs there
are far too many ways to decompose them into communities. But fast
approximate algorithms do exist \cite{Fortunato2010Community}, many based on the idea of
greedily merging small communities into larger ones with a higher modularity 
\cite{New:2004a,ClaNewMoo:2004}.  

For bipartite graphs the null model must be modified, to reproduce the
characteristic form of bipartite adjacency matrices 
\begin{equation*}
A=\left ( 
\begin{array}{cc}
0 & B \\ 
B^{T} & 0%
\end{array}
\right )
\end{equation*}
also for the null model \cite{Bar:2007}. This gives a bipartite modularity $Q_{B}$.
Comparatively few algorithms have been proposed for maximizing $Q_{B}$, but
methods for unipartite networks can often be adapted with little trouble.

Recently, Barber \cite{Bar:2007} proposed an appropriate algorithm
(BRIM: bipartite, recursively induced
modules) to find communities for bipartite networks. Starting from a (more
or less) ad hoc partition of the vertices of type 1, it is straightforward to
optimize a corresponding decomposition of the vertices of type 2. From
there, optimize the decomposition of vertices of type 1, and iterate. In
this fashion, modularity will increase until a (local) maximum is reached.
NetzCope allows to combine a suitable greedy algorithm with BRIM,
significantly enhancing the performance of the former.

Modularity has some limitations that should be kept in mind. The measure has
a resolution limit dependent on the number of edges in the network, so that
small communities cannot be found in large networks by simply finding a
maximum in the modularity\cite{ForBar:2007}. Further, modularity maximization is an
NP-complete problem\cite{Brandes2008}; typical to the class, there are exponentially many
local maxima in $Q$, so some ambiguity of decompositions is inevitable\cite{Good2010Performance}.
NetzCope provides a number of visualization tools to vary and compare them.
For a quantitative comparison, NetzCope will compute the mutual information 
\cite{DanDiaDucAre:2005}
of different decompositions.

\section{What NetzCope does}

For moderately large networks of some $10^{4}$ vertices, say, there is not
only the challenge of finding a display which exhibits as much as possible
of the network structure. There are simply too many vertices and edges to
fit distinguishably into a plot of any reasonable size.

As a consequence, a central part of our strategy will be to identify,
display, and analyze communities  within the overall network. 
These (interconnected) communities do admit a graphical representation, and so
NetzCope first displays a network of communities. For more detailed analysis
the software then allows the user to ``zoom
into'' communities  and  explore their inner
structure.

In contrast to well established network analysis software tools such as
UCINET or Pajek, Netzcope implements new methods to analyze and visualize
network structures, with a special emphasis on using recent methods from
statistical physics to identify and visualize community groups and to
analyze and visualize the adjacency matrix.

Overall, the principal functionalities of NetzCope are:
\begin{enumerate}
\item Identification of disjoint components within the network, if any such
exist. For any such component it will perform the following tasks:

\item Display of the adjacency matrix which encodes the connectivity

\item Display the same for the matrix after ``Fiedler
ordering'' which reorders the original (e.g., alphabetical)
network listing in such a way that interacting partners are grouped together

\item Generate plots of the nodes and the links between them in two or three dimensions, and rotate the plots about their axes 

\item Identify \emph{communities} of close
collaboration inside the network

\item Plot the network of these communities. This achieves one of the main
goals namely a suitable complexity reduction, so as to arrive at a feasible
and meaningful graphical display of networks of a size where plotting of the
full network becomes meaningless. The following items are tools for the
further analysis of these communities, such as

\item Topical profile: nodes may carry one or more labels. Their frequency
within a community is represented by colored segments in the aforementioned
plot. A more precise representation of this as a histogram can be called up
for each community, comparing their occurrence within the community to
overall occurrence in the network.

\item A scatter plot displays the number of internal versus external links
for the leading players in the community, thus identifying central players
and gatekeepers. Centrality is further analyzed by scatter plots which
compare \emph{high linkage} (to many partners) and \emph{important linkage}
(to important partners) for leading players.

\item An important functionality of NetzCope is the possibility to repeat
the above tasks for each  of the communities. The user may thus analyze
the network iteratively, ``zooming into'' the community structure until a
desired level of structural detail is obtained.

\item Finally ``network portraits'' as proposed by
Bagrow et al. \cite{Bagrow2008Portraits} are
included, mainly to facilitate the comparison of empirical network structure with that
of simulations stemming e.g. from random graph or multi-agent models.
\end{enumerate}
Originally designed to analyze and display bipartite networks of
organizations collaborating through projects, we have lifted this restriction
to make NetzCope more versatile. NetzCope can now also analyze general,
unipartite networks.

NetzCope distributions, together with sample network data files,
are available for Linux as well as Windows users. NetzCope loads network data in
the widely used Pajek format or from an edge list in plain text format. Netzcope can 
be downloaded at \url{http://www.physik.uni-bielefeld.de/~strogan/}.

\section{Some NetzCope Screenshots}

\subsection{The Connected Components}

\begin{figure}
	\includegraphics[width=\textwidth]{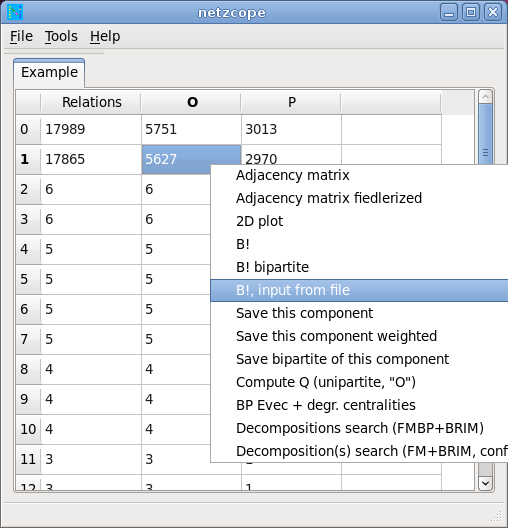}
	\caption{Connected Components}
	\label{fig:start}
\end{figure}

After a network has been loaded, NetzCope first extracts and lists the
connected components. For a bipartite network, each line in Fig.~\ref{fig:start} 
shows the number of edges (``relations''), the number of nodes of
type O (``organizations''), the number of
nodes of type P (``projects'').
Clicking on one of the boxes displayed will open a menu of options for the
further analysis of the chosen component.

\subsection{The Adjacency Matrix}

\begin{figure}
	\subfigure[Ad Hoc Ordering] {
	\includegraphics[width=.8\textwidth]{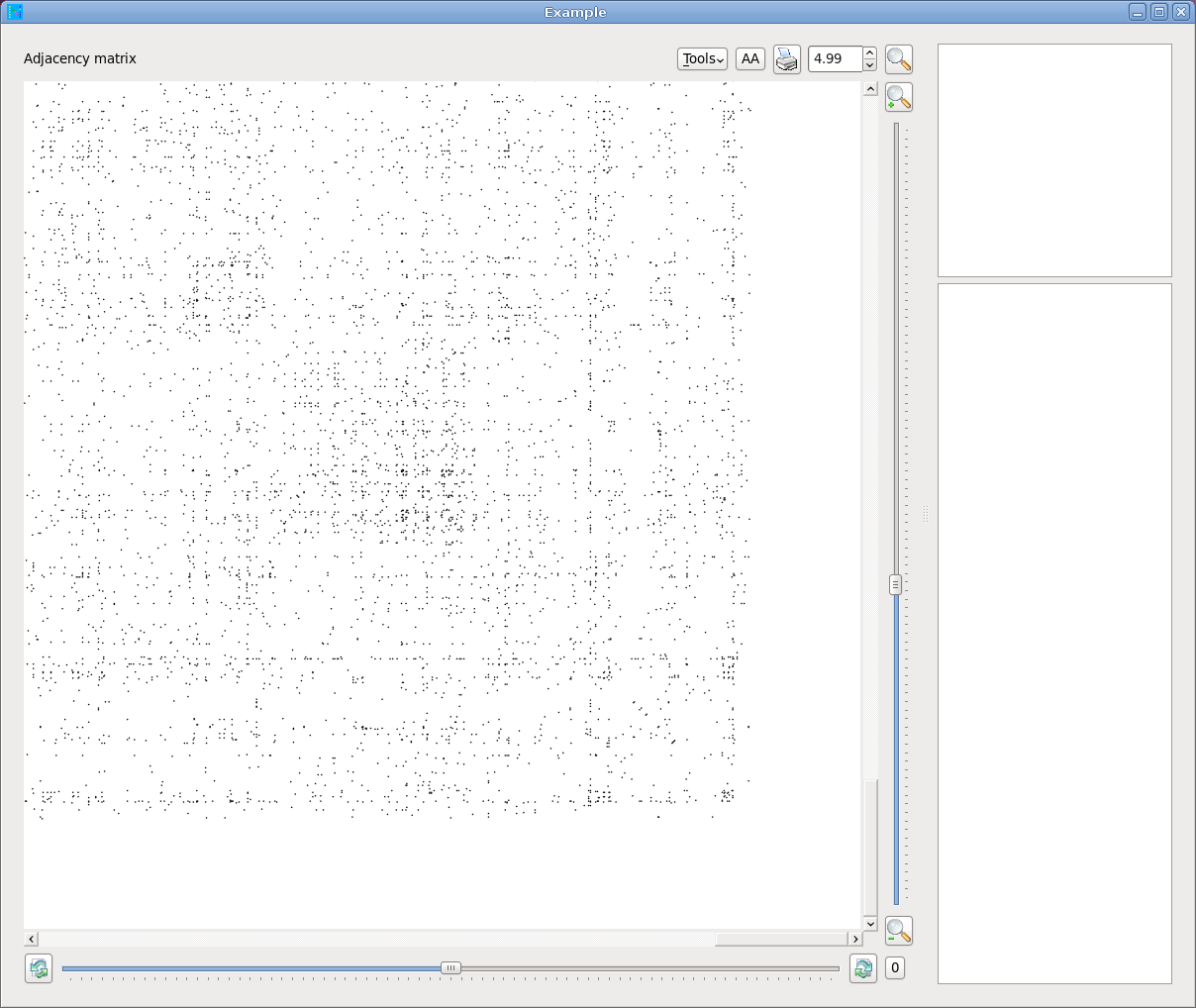}
	\label{fig:ad}
	}
	\subfigure[Fielder Ordering] {
	\includegraphics[width=.8\textwidth]{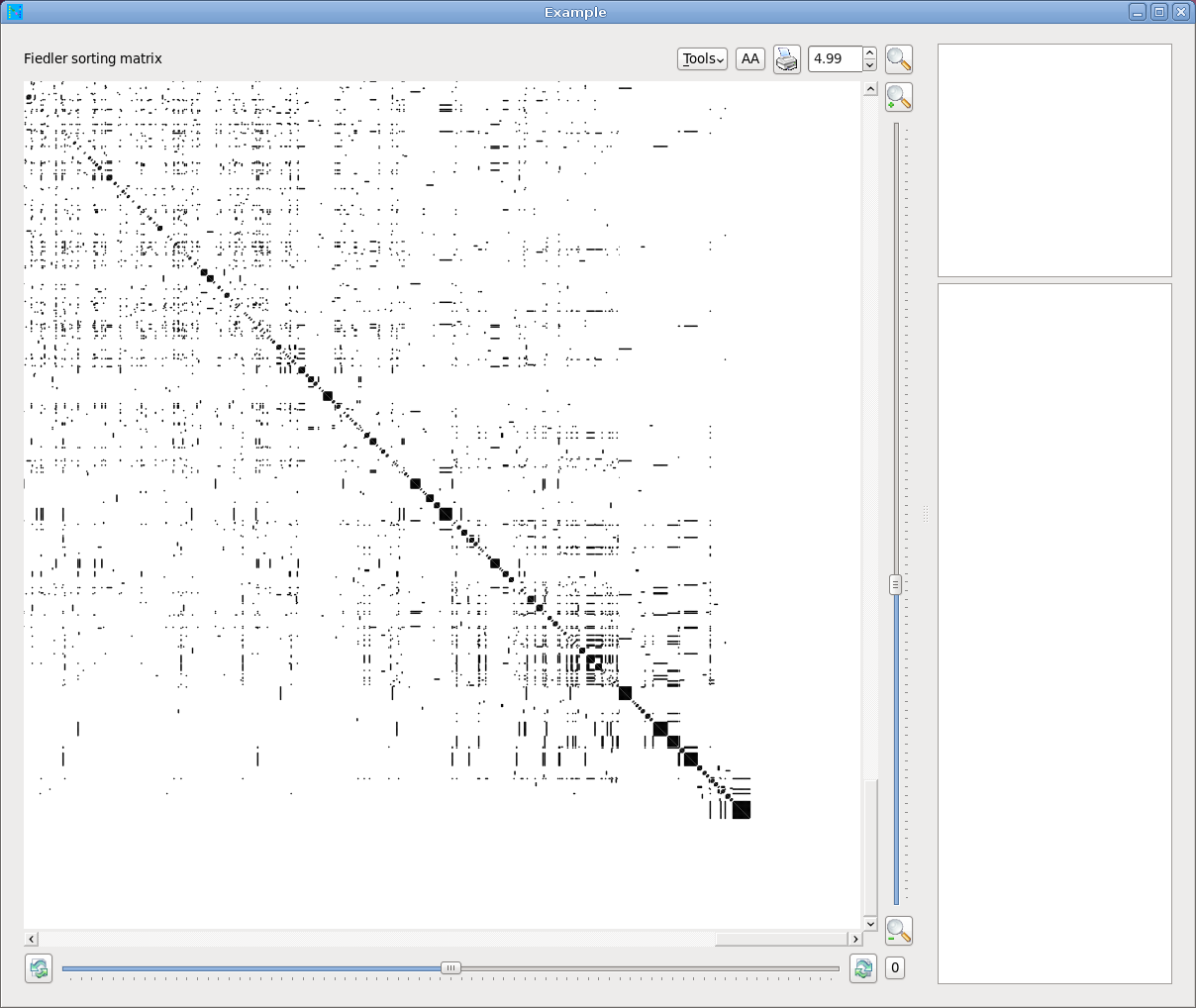}
	\label{fig:fied}
	}
	\caption{Adjacency Matrix}
\end{figure}

For large networks, NetzCope displays the non-zero entries (edges) of the
adjacency matrix as dots. Figure~\ref{fig:ad} 
shows part of such a matrix for
some 5600 vertices. They will in general be more or less randomly
distributed, as long as the list of vertices does not group vertices
together which are connected by many links. 

Fiedler ordering, reordering the graph vertices so that the components of the
Fielder vector are sorted, does just this.
Figure~\ref{fig:fied} 
shows the same segment of
the adjacency matrix after Fiedler ordering with its characteristic
concentration of dots near the diagonal, i.e., of short links in the listing.

%
%

\subsection{Plotting the Graph}

\begin{figure}
	\subfigure[The 2-d Plot] {
	\includegraphics[width=.8\textwidth]{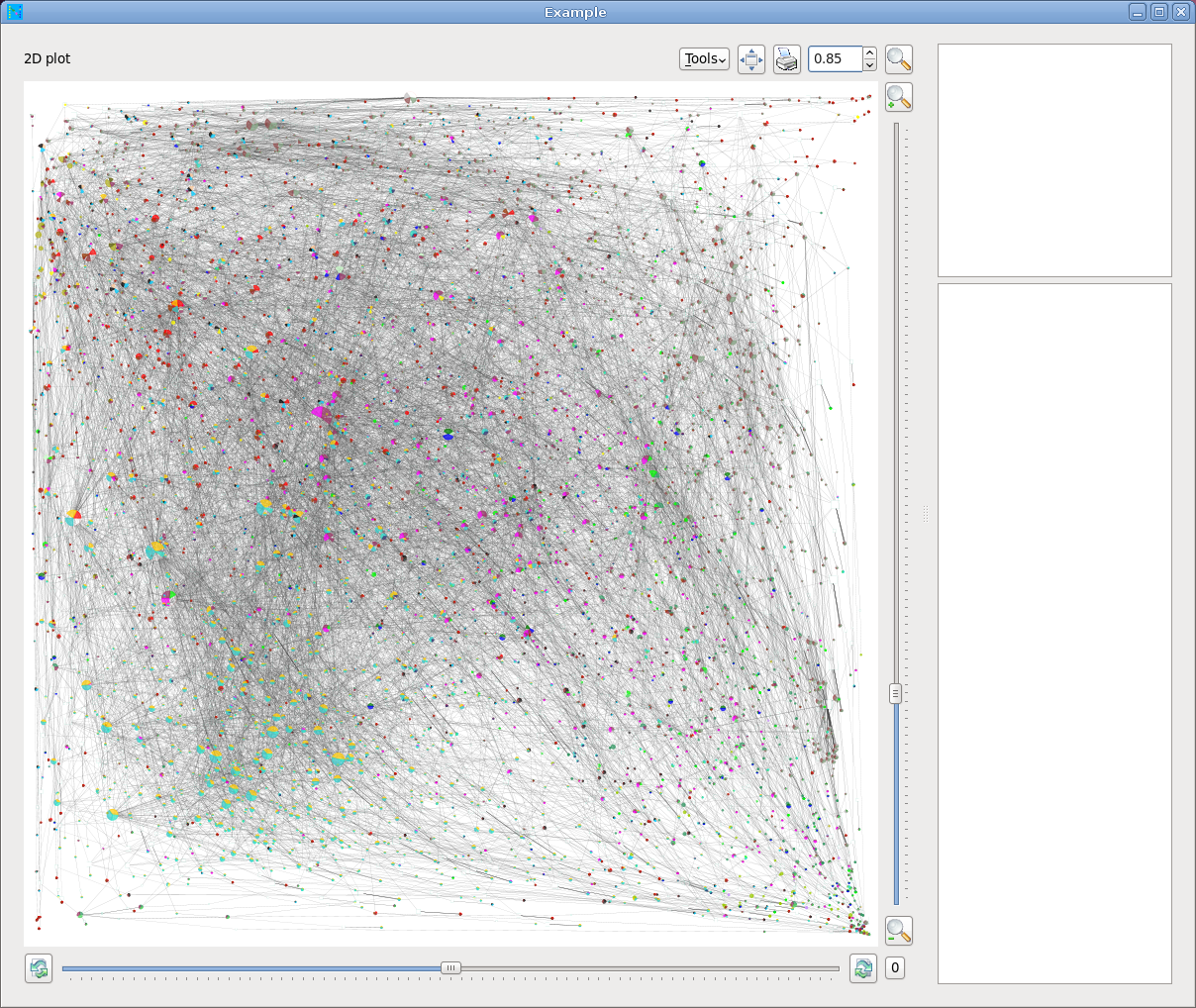}
	\label{fig:2d}
	}
	\subfigure[Zooming Into the Plot] {
	\includegraphics[width=.8\textwidth]{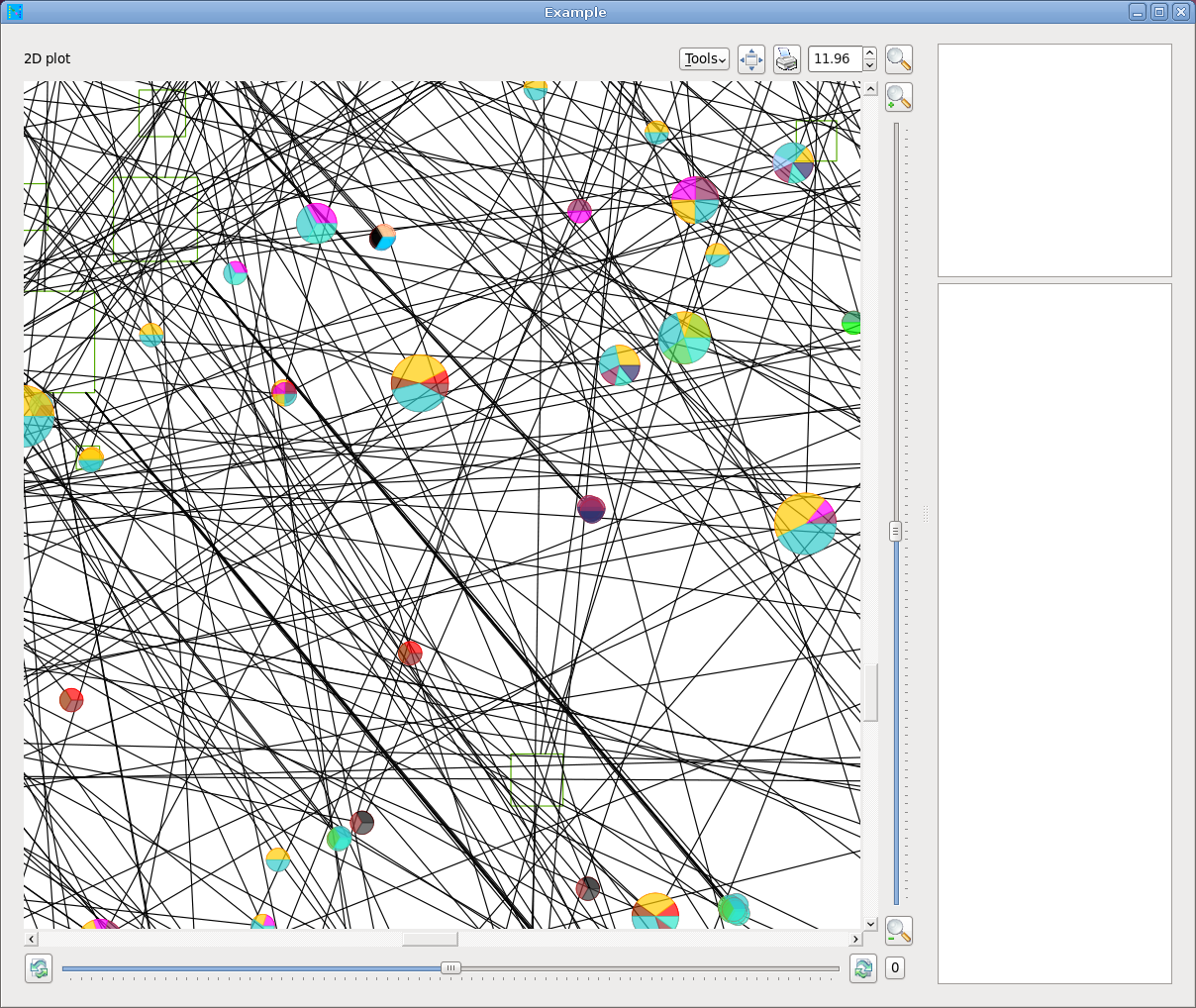}
	\label{fig:2dzoom}
	}
	\caption{Plotting the Graph}
\end{figure}

The menu item ``2 D plot'' will produce a
two dimensional representation of the graph. When networks have more than a
thousand vertices, the display will become more and more opaque (Fig.~\ref{fig:2d}),
and even the NetzCope zoom function will
be of limited use (Fig.~\ref{fig:2dzoom}).

\subsection{Plotting the Network of Communities}

\begin{figure}
	\includegraphics[width=\textwidth]{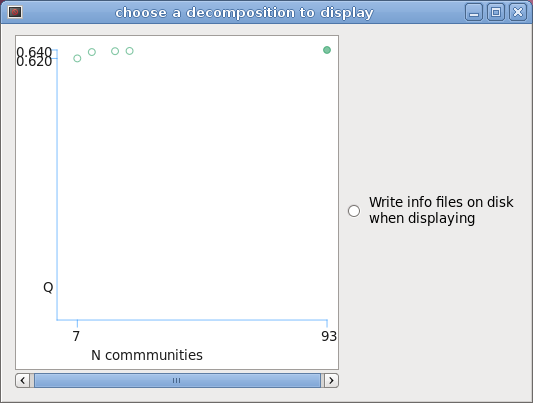}
	\caption{Modularities vs.~Number of Communities}
	\label{fig:Modularities}
\end{figure}

\begin{figure}
	\includegraphics[width=\textwidth]{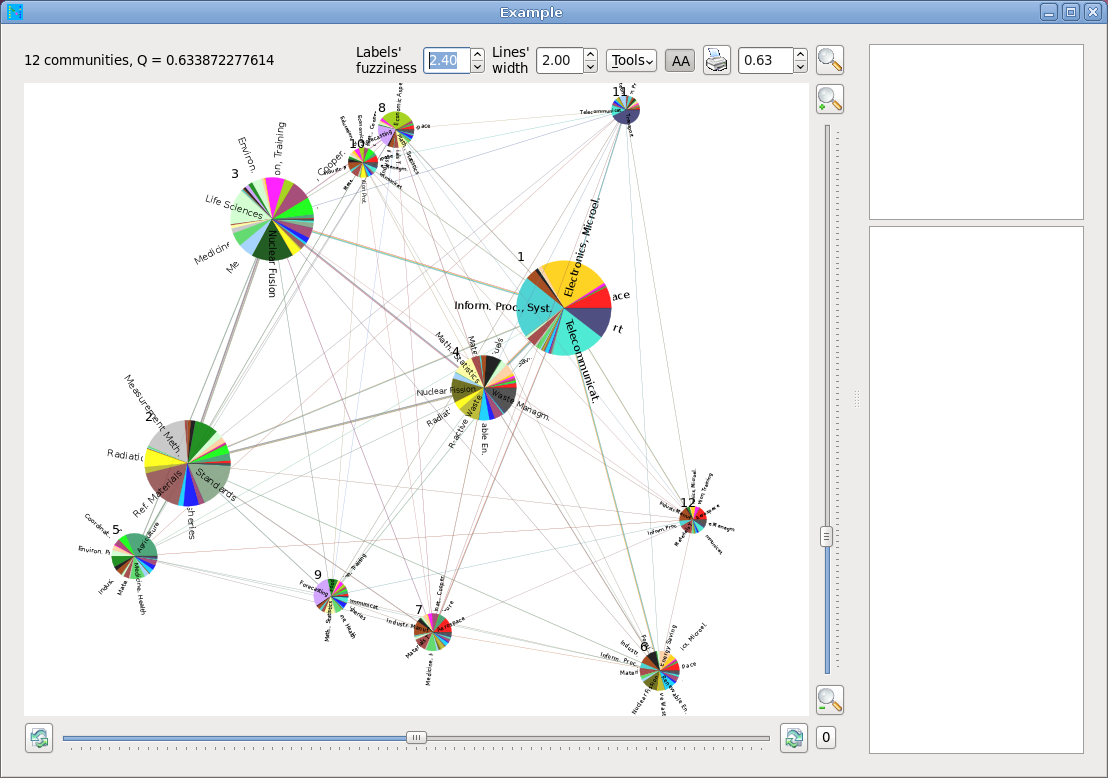}
	\caption{Network of Communities}
	\label{fig:communities}
\end{figure}

Before NetzCope outputs a graph of communities, various settings can be
specified, in particular the number(s) of communities for which NetzCope
then computes the decomposition, the respective modularities are displayed
as in Fig.~\ref{fig:Modularities}. 
For the collaborative research
networks one observes a typical leveling off of the modularity $Q$ as the
number of communities increases.
A click on the chosen value point on the display opens a window to prescribe
some graphical settings, and then produces a display of the corresponding
network of communities, such as in Fig.~\ref{fig:communities}.

\subsection{The Network Portrait}

\begin{figure}
	\includegraphics[width=\textwidth]{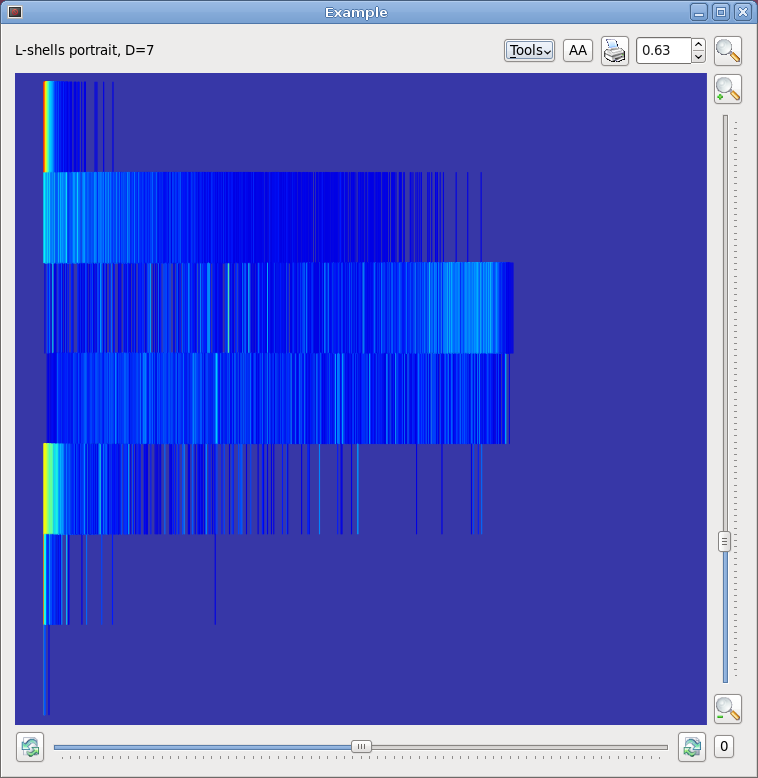}
	\caption{Network Portrait}
	\label{fig:portrait}
\end{figure}

The ``network portraits'' proposed by
Bagrow et al. \cite{Bagrow2008Portraits} display (Fig.~\ref{fig:portrait}) 
not only the degree distribution
but more generally, row by row, the distribution of all shell sizes 
where the $n$-th shell of a node in the network consists of all the
nodes at distance $n$---the first row gives the usual degree distribution,
while the highest $n$ is equal to the diameter of the network. These portraits
display structural features, including the network diameter ($d=7$ in our
example), which, particularly for large networks, are not accessible to
direct visual inspection of the latter. They have proven useful to show how
real world networks differ from certain random or multi-agent models.

\section{Auxiliary Features}

NetzCope provides numerous auxiliary features and settings, some of which
have already been mentioned in their proper context.
As mentioned above NetzCope allows interactive exploration of the network, 
permitting pointing into the display and reading
out the corresponding information at the side.
The display has a zooming and rotating capability, 
both in two and three dimensions. The latter is useful to
show possible congruences; network graphs occasionally become alike after a
suitable rotation.
Graphical details such as colorings and line widths can be chosen.

The search for communities can be done by maximizing modularity with 
a greedy algorithm \cite{New:2004a,ClaNewMoo:2004}
and/or BRIM \cite{Bar:2007} (for bipartites), and with varying degrees of
randomness and preferred numbers of communities.
Their display is controlled by the aforementioned Fiedler vector providing
one coordinate of the network nodes, and typically the subsequent
eigenvector of the normalized Laplacian to provide the second one. 
NetzCope allows the user to choose other eigenvectors for this purpose,
generating different distributions of the nodes within the display.

Starting from a bipartite network of ``organizations'' and ``projects,'' 
NetzCope can generate, e.g., the projected network
of organizations only, which are
linked if they share projects. The number of projects may additionally
be taken into account  to produce a \emph{weighted} graph.

One should also keep in mind that the community search may involve random
elements. In that case output will then vary from one run to another.
NetzCope offers a quantitative control of these variants by calculating
their mutual information.

\section{Conclusion}

We have presented a tour of the features of NetzCope, a software
application for the display and analysis of complex networks. 
Originally created to support investigating specific collaboration
networks, NetzCope has become a general purpose tool for network study.
NetzCope allows the user to interactively explore networks, especially
with regard to organizing the vertices so that the most important
relationships in the network can be observed. Despite its lengthy list
of features, NetzCope remains in its infancy, offering much potential
for future extension.

\section*{Acknowledgments} 

Work supported by FCT and POCI 2010 (project MAT/58321/2004)
with participation of FEDER, and by the Austrian Science Fund (FWF) under project P21450.

\bibliographystyle{ws-procs9x6}
\bibliography{qbicws}

\end{document}